\def\year{2019}\relax
\documentclass[letterpaper]{article} %DO NOT CHANGE THIS
\usepackage{aaai19}  %Required
\usepackage{times}  %Required
\usepackage{helvet}  %Required
\usepackage{courier}  %Required
\usepackage{url}  %Required
\usepackage{graphicx}  %Required
\usepackage{amsmath}
\begin{document}
\title{Symmetrization for Embedding Directed Graphs}
\author{\small Jiankai Sun \and  Srinivasan Parthasarathy \\
Department of Computer Science and Engineering, The Ohio State University\\
Email: sun.1306@osu.edu, srini@cse.ohio-state.edu
}
\maketitle
%\begin{abstract}
%AAAI creates proceedings, working notes, and technical reports directly from electronic source furnished by the authors. To ensure that all papers in the publication have a uniform appearance, authors must adhere to the following instructions. 
%\end{abstract}

%\noindent Congratulations on having a paper selected for inclusion in an AAAI Press proceedings or technical report! This document details the requirements necessary to get your accepted paper published using \LaTeX{}. If you are using Microsoft Word, instructions are provided in a different document. If you want to use some other formatting software, you must obtain permission from AAAI Press first. 

% \input{tex/abstract.tex}

Recently, one has seen a surge of interest in developing such methods including ones for learning such representations for (undirected) graphs (while preserving important properties)~\cite{SEANO}.
%The problem of graph embedding seeks to represent vertices of a graph in a low-dimensional vector space in which meaningful semantic, relational and structural information conveyed by the graph can be accurately captured ~\cite{cao2015grarep,Ma2018WSDM}. Classic vector-based machine learning techniques can leverage such vectors as features for many tasks such as link prediction, multi-label classification, clustering and vertex recommendation ~\cite{Harp2017,Wang2018WSDM}.
However, most of the work to date on embedding graphs has targeted undirected networks and very little has focused on the thorny issue of embedding directed networks. In this paper, we instead propose to solve the directed graph embedding problem via a two-stage approach: in the first stage, the graph is symmetrized in one of several possible ways, and in the second stage, the so-obtained symmetrized graph is embedded using any state-of-the-art (undirected) graph embedding algorithm.  Note that it is not the objective of this paper to propose a new (undirected) graph embedding algorithm or discuss the strengths and weaknesses of existing ones; all we are saying is that whichever be the suitable graph embedding algorithm, it will fit in the above proposed symmetrization framework. 

Satuluri et al. proposed various ways (such as Bibliometric and Degree-discounted symmetrization) of symmetrizing a directed graph into an undirected graph, while information about directionality is incorporated via weights on the edges of the transformed graph (or applying a re-weighting scheme in case of already weighted graphs)~\cite{Satuluri2011}. 

\begin{figure}[ht!]
    \small
    \centering
    \includegraphics[width=0.3\textwidth]{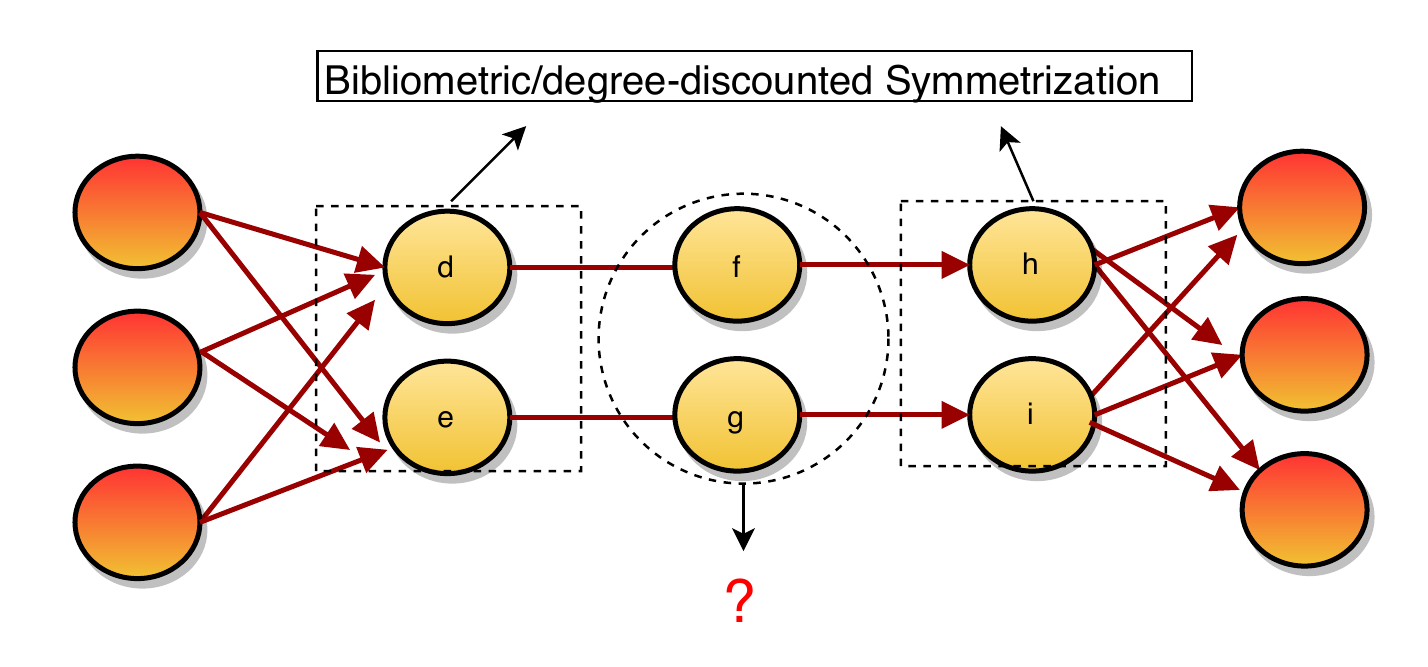}
    \caption{%Bibliometric and degree-discounted symmetrization can make nodes $d$ and $e$ (and also nodes $h$ and $i$) form a natural cluster since $d$ and $e$ are pointed to by the same nodes (for $h$ and $i$, they point to the same nodes). The nodes $f$ and $g$ neither point to the same nodes nor they are pointed to by the same nodes. Hence, 
    Bibliometric and degree-discounted symmetrization will not connect $f$ and $g$ in the resulting undirected graph. However, $f$ and $g$ should form a natural cluster as they have close common successors and predecessors.}
    \label{fig:graph_symmetrization_reachability}
\end{figure}
 
However, both bibliometric and degree-discounted symmetrization only consider first order graph structure while failing to take higher order graph structure into account. As shown in Figure~\ref{fig:graph_symmetrization_reachability}, bibliometric and degree-discounted symmetrization will not connect $f$ and $g$ in the resulting undirected graph. However, $f$ and $g$ should be connected and form a natural cluster as they can reach the same nodes and are also reached by the same nodes. We refer to the nodes which can reach a target node $i$ as $i$'s predecessors, and the nodes which can be reached by $i$ as $i$'s successors. Transitive closure (TC) of a directed graph is a methodology (usually housed in a simple data structure) that makes it possible to answer reachability questions. Let $G$ be the initial directed graph with adjacency matrix $\boldsymbol{A}$. The TC of $G$ is a graph $G^{+} = (V,E^{+})$ such that for all $v$, $w$ in $V$ there is an edge $(v,w)$ in $E^{+}$ if and only if there is a non-null path from $v$ to $w$ in $G$. 
The adjacency matrix of $G^{+}$ is represented as $\boldsymbol{A}_{G^+}$. Similar to out-link similarity $\boldsymbol{B}_{out}$ defined by Satuluri et al.~\cite{Satuluri2011}, we define out-reach similarity $\boldsymbol{B}_{o}$ as:

\begin{equation}
    \boldsymbol{B}_{o}(i,j) = \frac{1}{{(\boldsymbol{D}_{o}^{G^+}(i))}^{\alpha}{(\boldsymbol{D}_{o}^{G^+}(j))}^{\alpha}}\sum_k\frac{\boldsymbol{A}_{G^+}(i,k)\boldsymbol{A}_{G^+}^{T}(k,j)}{{(\boldsymbol{D}_{i}^{G^+}(k))}^{\beta}}
\end{equation}

where $\boldsymbol{D}_{o}^{G^+}$ is the diagonal matrix of out-degrees in $G^{+}$, $\boldsymbol{D}_{i}^{G^+}$ is the diagonal matrix of in-degrees
in $G^{+}$, $\alpha$ and $\beta$ are the discounting parameters. $\alpha = \beta = 0.5$ is found to work the best empirically. The above expression is symmetric in $i$ and $j$. And $\boldsymbol{B}_{o}$ is represented as:

\begin{equation}
        \boldsymbol{B}_{o} = {(\boldsymbol{D}_{o}^{G^+})}^{-\alpha}\boldsymbol{A}_{G^+}{(\boldsymbol{D}_{i}^{G^+})}^{-\beta}\boldsymbol{A}_{G^+}^T{(\boldsymbol{D}_{o}^{G^+})}^{-\alpha}
\end{equation}
And in-reach similarity $\boldsymbol{C}_{i}$ is defined as:

\begin{equation}
    \boldsymbol{C}_{i} = {(\boldsymbol{D}_{i}^{G^+})}^{-\beta}\boldsymbol{A}_{G^+}^T{(\boldsymbol{D}_{o}^{G^+})}^{-\alpha} \boldsymbol{A}_{G^+}{(\boldsymbol{D}_{i}^{G^+})}^{-\beta}
\end{equation}

 The resulting symmetrized undirected graph is represented as $G_U$. and its associated adjacency matrix is $\boldsymbol{A}_U$ which is the sum of $\boldsymbol{B}_o$ and $\boldsymbol{C}_{i}$:
\begin{equation}
\begin{aligned}
    \boldsymbol{A}_U = \boldsymbol{B}_{o} + \boldsymbol{C}_{i} 
    \end{aligned}
\end{equation}

\begin{figure}[h]
    \small
    \centering
    \includegraphics[width=0.3\textwidth]{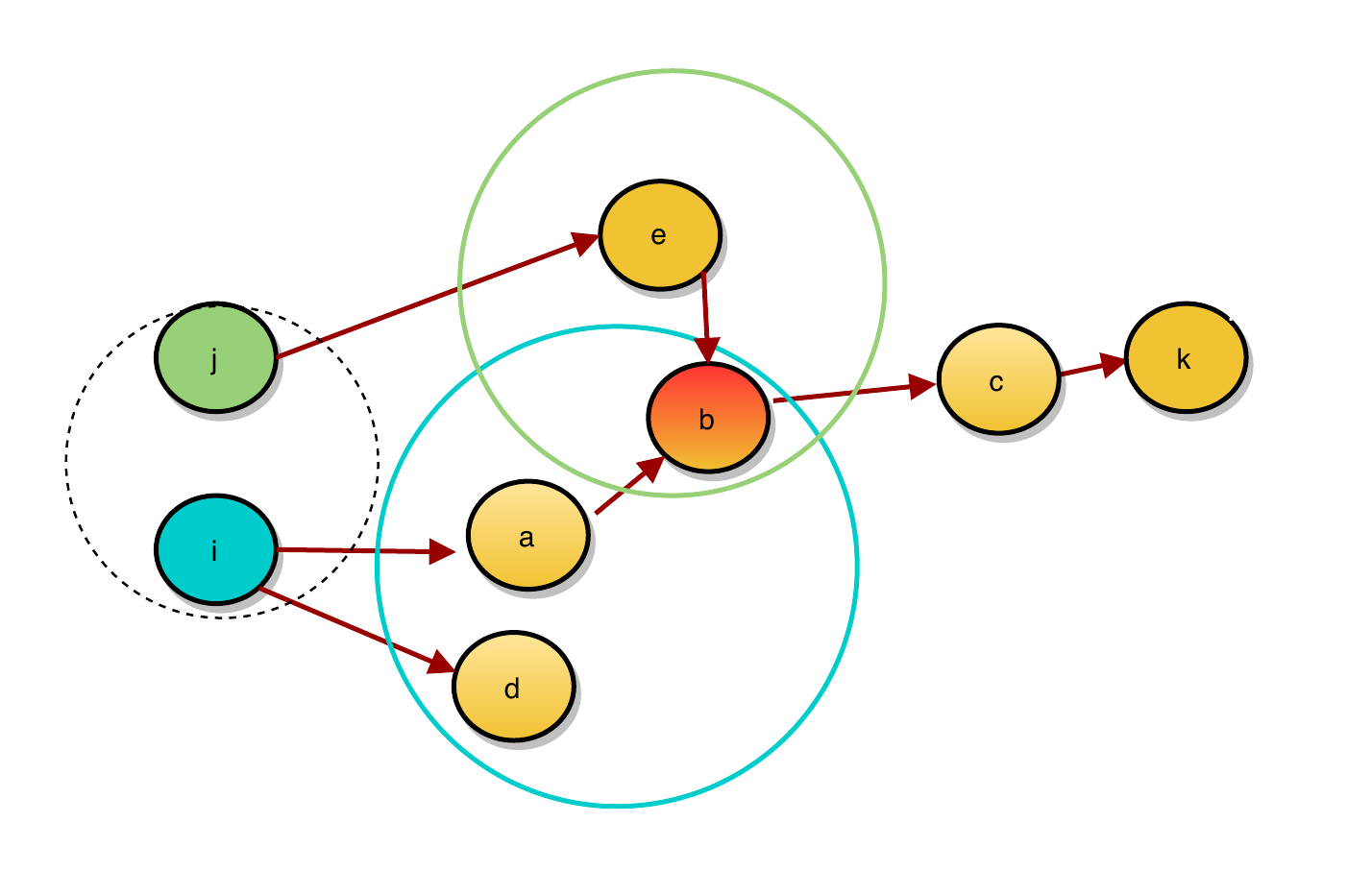}
    \caption{Illustration of Local Transitive Closure: $i$ can reach $a$, $b$, and $d$ in $l = 2$ steps. $j$ can reach $b$ and $e$ in $l=2$ steps. $b$ will contribute to the similarity of $i$ and $j$, while $c$ and $k$ will not.}
    \label{fig:graph_symmetrization_LTC}
    % \vspace{-0.2in}
\end{figure}

 However, computing TC for large directed graphs with cycles is expensive. Instead, we propose to do a breadth first search (BFS) with a depth constraint $l$ to compute a node's local reachability, as shown in Figure \ref{fig:graph_symmetrization_LTC}. 
This approach generalizes Bibliometric and Degree-discounted. For example, if we set $l = 1$, the resulting undirected graph should be the same as Bibliometric and Degree-discounted symmetrization.  

We also include hierarchical difference between node pairs and distance to common successors and predecessors to refine $\boldsymbol{A}_U$ as follows.

% We also include two other intuition to refine $\boldsymbol{A}_U$ as follows:

% \begin{itemize}
%     \item Hierarchical difference between node pairs 
%     \item Distance to common successors and predecessors 
% \end{itemize}

\subsubsection{\textbf{Hierarchical difference between node pairs }}

\begin{figure}[h]
    \small
    \centering
    \includegraphics[width=0.35\textwidth]{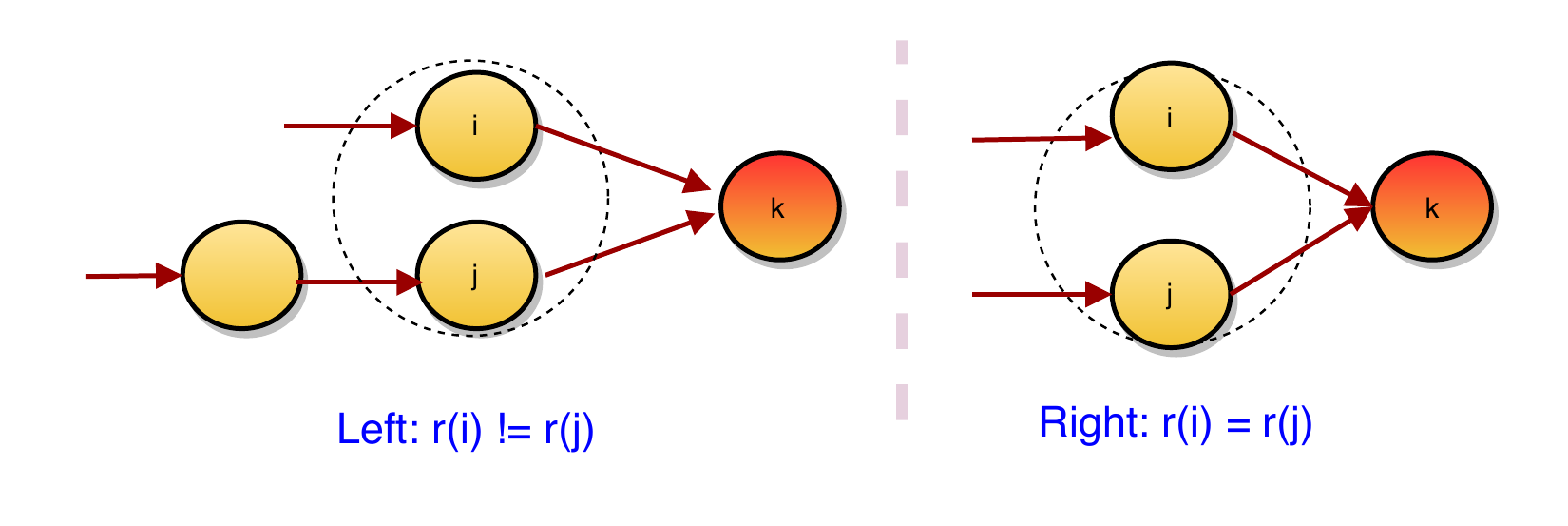}
    \caption{All else equal, the node $i$ should be less similar to the node $j$ which has a different hierarchy (\textbf{left}) when compared to the condition that both nodes $i$ and $j$ have the same hierarchical information in the graph (\textbf{right})}
    \label{fig:graph_symmetrization_hierarchyDifferences}
    % \vspace{-0.2in}
\end{figure}

As shown in Figure~\ref{fig:graph_symmetrization_hierarchyDifferences}, another intuition suggests that the out-reach similarity between $i$ and $j$ should be inversely related to the hierarchical difference between $i$ and $j$. The hierarchy score of each node can be assigned by a function to represent where it stands in the entire network~\cite{Sun2017}. A higher hierarchical difference contributes less to the in/out-reach similarity. 

\subsubsection{\textbf{Hierarchical distance to common successors and predecessors }}

\begin{figure}[ht]
    \small
    \centering
    \includegraphics[width=0.35\textwidth]{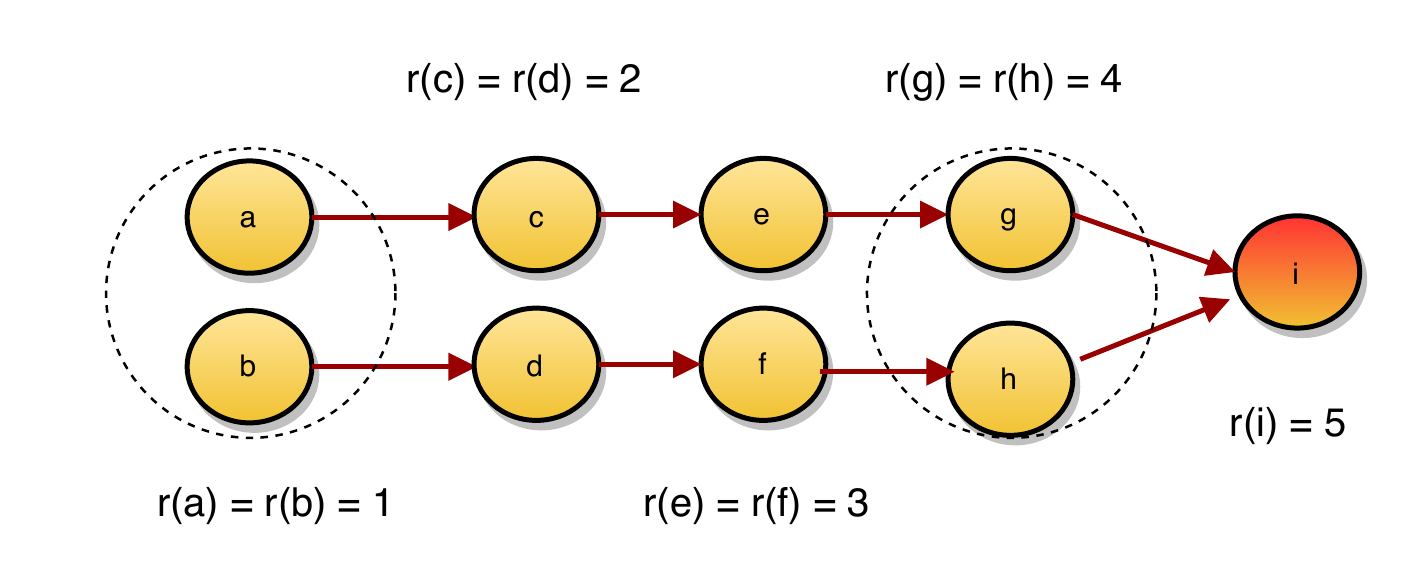}
    \caption{Even though nodes $a$, $b$, $g$, and $h$ can reach the same node $i$, the contribution of this event to the out-reach similarity between $a$ and $b$ is smaller than $g$ and $h$, as $g$ and $h$ are closer to $i$ than $a$ and $b$.}
    \label{fig:graph_symmetrization_path_length}
    % \vspace{-0.1in}
\end{figure}

As shown in Figure~\ref{fig:graph_symmetrization_path_length}, intuition suggests that when nodes $a$, $b$, $g$, and $h$ can reach the same node $i$, the contribution of this event to the out-reach similarity between $a$ and $b$ is smaller than $g$ and $h$, as $g$ and $h$ are closer to $i$ than $a$ and $b$. To leverage above intuition, we can use hierarchical difference to measure distance between node pairs. The out-reach similarity between node $i$ and $j$ should be inversely related to the hierarchical difference $\boldsymbol{M}_{i,k}$ and $\boldsymbol{M}_{j,k}$, where $k$ is a node which can be reached by $i$ and $j$, and $\boldsymbol{M}_{i,k}$ represents the hierarchical difference between node $i$ and $k$.

Generally, by using our proposed symmetrization method, node pairs, having close common successors and/or predecessors and sharing similar hierarchy information in the input directed graph, will be connected in the resulting undirected graph.

% \vspace{-0.1in}
\subsection{Application: Expert Finding in CQAs}
 Existing methods for expert finding in community question answering services (CQAs)~\cite{QDEE2018,ColdRoute} suffer from the data sparseness problem. In this paper, we propose to symmetrize the directed CQA graphs to undirected graphs, which can add more interactions in the corresponding undirected graphs to overcome the data sparseness problem. We then train a deep neural network based regressor with input as $<\boldsymbol{X}, \boldsymbol{Y}>$, where each input $\boldsymbol{x}$ concatenates the feature vector of a question $q$ and its corresponding answerer $u$, and $y$ is the voting score of the answer  provided for $q$ by $u$, since Sun et al. discovered that voting score can be viewed as an indicator to identify the best answerer~\cite{ColdRoute}. The feature vectors of questions and answerers can be obtained by applying any suitable (undirected) graph embedding method to the generated undirected graphs. Given a new question $q$ and a set of potential answerers $C_q$, we can predict each candidate $u$'s voting score for $q$, where $u \in C_q$. The user who achieves the highest voting score will be selected as the best answerer for $q$.

 {\bf Acknowledgments} This work is supported by NSF grants CCF-1645599, IIS-1550302, and CNS-1513120, RI xxxxxx, and a grant from the Ohio Supercomputer Center (PAS0166). All content represents the opinion of the authors, which is not necessarily shared or endorsed by their sponsors.

\bibliographystyle{aaai}
\bibliography{sigproc}

\begin{thebibliography}{}

\bibitem[\protect\citeauthoryear{Liang \bgroup et al\mbox.\egroup
  }{2018}]{SEANO}
Liang, J.; Jacobs, P.; Sun, J.; and Parthasarathy, S.
\newblock 2018.
\newblock Semi-supervised embedding in attributed networks with outliers.
\newblock In {\em SIAM SDM}.

\bibitem[\protect\citeauthoryear{Satuluri and
  Parthasarathy}{2011}]{Satuluri2011}
Satuluri, V., and Parthasarathy, S.
\newblock 2011.
\newblock Symmetrizations for clustering directed graphs.
\newblock In {\em EDBT/ICDT}.

\bibitem[\protect\citeauthoryear{Sun \bgroup et al\mbox.\egroup
  }{2017}]{Sun2017}
Sun, J.; Ajwani, D.; Nicholson, P.~K.; Sala, A.; and Parthasarathy, S.
\newblock 2017.
\newblock Breaking cycles in noisy hierarchies.
\newblock In {\em Proceedings of the 2017 ACM on Web Science Conference},
  WebSci '17,  151--160.

\bibitem[\protect\citeauthoryear{{Sun} \bgroup et al\mbox.\egroup
  }{2018a}]{QDEE2018}
{Sun}, J.; {Moosavi}, S.; {Ramnath}, R.; and {Parthasarathy}, S.
\newblock 2018a.
\newblock {QDEE: Question Difficulty and Expertise Estimation in Community
  Question Answering Sites}.
\newblock In {\em ICWSM}.

\bibitem[\protect\citeauthoryear{Sun \bgroup et al\mbox.\egroup
  }{2018b}]{ColdRoute}
Sun, J.; Vishnu, A.; Chakrabarti, A.; Siegel, C.; and Parthasarathy, S.
\newblock 2018b.
\newblock Coldroute: effective routing of cold questions in stack exchange
  sites.
\newblock In {\em ECML PKDD}.

\end{thebibliography}

\end{document}